\documentclass[aps,floats,twocolumn,nofootinbib,preprintnumbers]{revtex4}

\usepackage{amssymb}
\usepackage{amsmath}

\usepackage{graphicx}
\usepackage{graphics}
\usepackage{dcolumn}
\usepackage{color}
\usepackage{rotate}
\usepackage{fancyhdr}
\usepackage{hyperref}
\usepackage{indentfirst}
\usepackage{multirow}
\usepackage{umoline}
\usepackage{hyperref}

\def\bea{\begin{eqnarray}}
\def\eea{\end{eqnarray}}

\begin{document}

\title{Supersymmetry versus Compositeness: 2HDMs tell the story}

\author{Stefania De Curtis$^a$}
\author{Luigi Delle Rose$^{a}$}
\author{Stefano Moretti$^{b,c}$} 
\author{Kei Yagyu$^a$}

\affiliation{
$^a$INFN, Sezione di Firenze, and Department of Physics and Astronomy, University of Florence, Via G. Sansone 1, 50019 Sesto Fiorentino, Italy\\
$^b$School of Physics and Astronomy, University of Southampton, Highfield, Southampton SO17 1BJ, United Kingdom\\
$^c$Particle Physics Department, Rutherford Appleton Laboratory, Chilton, Didcot, Oxon OX11 0QX, United Kingdom
}


\begin{abstract}
\noindent
Supersymmetry and Compositeness are two prevalent paradigms providing both a solution to the hierarchy problem and
a motivation for a light Higgs boson state. As the latter has now been found, its dynamics can hold the key to disentangle the two theories. 
An open door towards the solution is found in the context of 2-Higgs Doublet Models (2HDMs), which are necessary to 
Supersymmetry and natural within Compositeness in order to enable Electro-Weak Symmetry Breaking. 
We show how 2HDM spectra of masses and couplings accessible at the Large Hadron Collider may allow one to separate the two scenarios.
 
\end{abstract}

\maketitle

{\it Introduction} --
What has been the discovery of the cornerstone of the Standard Model (SM), the Higgs boson, should now turn out to be the stepping stone into a new physics world.  The latter is currently unknown to us. We know though what 
 its dynamics should prevent,  i.e., the hierarchy problem.
There are two possible pathways to follow in the quest for Beyond the SM (BSM) physics which would at once reconcile the above experimental and theoretical instances. These are Supersymmetry and Compositeness.  The former predicts a light Higgs boson as it relates its trilinear self-coupling (hence its mass) to the smallness of the gauge couplings and avoids the hierarchy problem thanks to the presence of supersymmetric counterparts, of different statistics with respect to the SM objects (a boson for a fermion, and vice versa), which then cancel divergent contributions to the Higgs mass. In fact, both conditions are achieved most effectively when the top quark partner in Supersymmetry, the stop, has a mass above the
TeV scale, thereby lifting the tree-level Higgs boson mass from below $M_Z$ to the measured value of 125 GeV or so.
The latter can also naturally embed a light Higgs state, as a pseudo Nambu-Goldstone Boson (pNGB),  if the top quark composite counterparts are at the TeV scale.

It is thus evident that, despite the different principles behind the two theories,  the low energy phenomenology of their Higgs and  top partner sectors  may be rather similar. However, in this connection, we shall prove here that differences exist between these two BSM scenarios that can presently be tested at the Large Hadron Collider (LHC). This can become manifest if one realises that an enlarged Higgs sector, notably involving a 2-Higgs Doublet Model (2HDM) dynamics, is required by Supersymmetry and natural in Compositeness. By exploiting a recently completed calculation of the Higgs potential \cite{BigPaper} in a Composite 2HDM (C2HDM), we will contrast the ensuing results concerning Higgs boson masses and couplings to the well known ones established in Supersymmetry \cite{Djouadi}. 

{\it Explicit Model} --
We now give a brief setup for our C2HDM construction to describe the salient features of the numerical results shown below. Details will be illustrated in \cite{BigPaper}.
We consider a spontaneous global symmetry breaking $SO(6)\to SO(4)\times SO(2)$ at a (Compositeness) scale $f$ providing two pNGB doublets. 
The pNGB matrix $U$ is constructed from the 8 broken $SO(6)$ generators ($T^{\hat{a}}_{i}$, $i=1,2$, $\hat{a}=1,\dots,4$) out of the 15 total ones ($T^A$, $A=1,\dots, 15$) as
\begin{align}
&U=e^{i\frac{\Pi}{f}},\quad
\Pi\equiv \sqrt{2}\phi_{i}^{\hat{a}}T^{\hat{a}}_{i}=-i 
\begin{pmatrix}
0_{4\times 4} & {\bf \Phi} \\
-{\bf \Phi}^T & 0_{2\times 2}  
\end{pmatrix},   \label{u6}
\end{align}
where ${\bf \Phi} \equiv (\phi_1^{\hat{a}},\phi_2^{\hat{a}}) $. 
The two real 4--vectors $\phi_i^{\hat{a}}$ can be rearranged into two complex doublets $\Phi_i$ ($i=1,2$), as 
\begin{align}
\Phi_i = \frac{1}{\sqrt{2}}\left(
\begin{array}{cc}
\phi_i^{\hat 2} + i\phi_i^{\hat 1} \\
\phi_i^{\hat 4} - i\phi_i^{\hat 3} 
\end{array}\right). \label{doublet6}
\end{align}
The Vacuum Expectation Values (VEVs) of the Higgs fields 
are taken to be $\langle \phi_i ^{\hat 4}\rangle = v_i$, so we define $v^2 = v_1^2 + v_2^2$ and $\tan\beta = v_2/v_1$ as usual in  2HDMs. 
Because of the non-linear nature of the pNGBs eventually emerging as Higgs boson states, $v$ does not correspond to the SM Higgs VEV $v_{\text{SM}}^{}$ (related to the Fermi constant $G_F$ by 
$v_{\text{SM}}^2 = (\sqrt{2}G_F)^{-1}$), rather, it satisfies the relation  
\begin{align}
v_{\text{SM}}^2 = f^2\sin^2 \frac{v}{f}. 
\end{align}

In order to obtain a non-zero value for the Higgs boson masses, explicit breaking terms of the $SO(6)$ symmetry must be introduced.  
Within the partial Compositeness paradigm~\cite{partial}, this can be achieved by a linear mixing between the (elementary) SM  and (composite)  strong sector fields, where the former and the latter are described, respectively,
by $SU(2)_L \times U(1)_Y$ and $SO(6)\times U(1)_X$ representations. Notice that, in the strong sector, 
the $U(1)_X$ symmetry is necessary to correctly realise the $U(1)_Y$ hypercharge charge assignment of the SM fermions. 
Thus,  the Lagrangian is 
\begin{align}
{\cal L} = {\cal L}_{\text{SM}} + {\cal L}_{\text{strong}} + {\cal L}_{\text{mix}}, \label{lag}
\end{align}
where ${\cal L}_{\text{SM}}$ denotes the Lagrangian of kinetic terms for the SM gauge bosons ($W_\mu^a$ and $B_\mu$) and fermions 
(for the computation of the scalar potential it is sufficient to consider only the third generation of the left-handed quark doublets $q_L$ and right-handed top quark $t_R^{}$), while 
${\cal L}_{\text{strong}}$ and ${\cal L}_{\text{mix}}$ represent the Lagrangian of the strong sector and the mixing one, respectively.

For an explicit realisation of this setup, it is convenient to define the elementary fields as incomplete $SO(6)$ multiplets by introducing spurion fields. 
Namely, the elementary gauge bosons can be embedded into $SO(6)$ adjoint $(A_\mu^A)$ and $U(1)_X$ $(X_\mu)$ spurion fields, while
the elementary fermions can be embedded into the $SO(6)$ ${\bf 6}$-plet spurions 
$q_{L}^{\bf{6}}$   and $t_R^{\bf{6}}$.
 
The explicit model is  based on the 2-site construction defined in Ref.~\cite{4dchm}.
The extra degrees of freedom in the gauge  sector are the spin-1 resonances of the adjoint of $SO(6)$ ($\rho^A$) and  $U(1)_X$ ($\rho_X$).   
The gauge sector for ${\cal L}_{\text{strong}}$ and ${\cal L}_{\text{mix}}$ is then given by 
\begin{align}
&{\cal L}_{\text{strong}}^ {\text{gauge}}+ {\cal L}_{\text{mix}}^{\text{gauge}} = -\frac{1}{4g_\rho^2} \rho^A_{\mu\nu}\rho^{A\mu\nu}
-\frac{1}{4g_{\rho_X}^2}\rho_{X\mu\nu}\rho_X^{\mu\nu} \notag\\
&~~+\frac{f_1^2}{4}\text{tr}(D_\mu U_1)^\dagger (D_\mu U_1) +\frac{f_2^2}{4}\text{tr}(D_\mu \Sigma_2)^T(D^\mu \Sigma_2), \label{4dlag}
\end{align}
where $U_i = e^{i\frac{f}{f_i^2}\Pi}$, $U=U_1 U_2$ and $\Sigma_i = U_i \Sigma_0U_i^T~(i = 1,2)$, with 
$f^{-2} = f_1^{-2} + f_2^{-2}$ and 
 $\Sigma_0 = 0_{4\times 4}\oplus i\sigma_2$ being an $SO(4)\times SO(2)$ invariant vacuum. 
Also,
\begin{align}
&D_\mu U_1 = \partial_\mu U_1 -iA_\mu U_1 +i U_1 \rho_\mu, \notag\\
&D_\mu \Sigma_2 = \partial_\mu \Sigma_2 - i[\rho_\mu, \Sigma_2], 
\end{align}
where  $\rho_\mu \equiv \rho_\mu^A T^A+ \rho_\mu^X T^X$ and $A_\mu \equiv A_\mu^A T^A+ X_\mu T^X$.

In the fermion sector we consider $N$ spin $1/2$ resonances $\Psi^I$ ($I=1,\dots, N$) which are $SO(6)$ ${\bf 6}$-plets with $X = 2/3$. 
The corresponding Lagrangian is 
\begin{align}
&{\cal L}_{\text{strong}}^{\text{ferm}}  + {\cal L}_{\text{mix}}^{\text{ferm}}  = 
 \bar{\Psi}^I iD\hspace{-2.2mm}/\hspace{0.6mm} \Psi^I 
 - \bar{\Psi}_{L}^I M_\Psi^{IJ} \Psi_R^J  - \bar{\Psi}_L^I (Y_1^{IJ} \Sigma_2 \notag\\ & + {Y}_2^{IJ} \Sigma_2^2 )\Psi_R^J  + (\Delta_L^{I}\bar{q}_L^{\bf{6}} U_1\Psi_R^I +  \Delta_R^I \bar{t}_R^{\bf{6}} U_1\Psi_L^I ) + 
\text{h.c.}, \label{ferm}
\end{align}
where $\Delta_{L,R}$ ($M_\Psi$, $Y_1$ and ${Y}_2$) are dimensionful $N$--vectors ($N\times N$ matrices).
Since $\Sigma_2^3 = -\Sigma_2$, terms up to quadratic power in $\Sigma_2$ reproduce the most general interaction Lagrangian between the fermionic resonances and the pNGB fields. 
For simplicity, we assume CP-conservation in the strong sector, i.e., all parameters in Eq.~(\ref{ferm}) to be real. 
As a result, the Coleman-Weinberg (CW) Higgs potential is CP-conserving. The Yukawa interactions 
for the right-handed bottom quark can also be included by introducing further spin-1/2 resonances with $X=-1/3$.  
Analogously one can implement  partial Compositeness for tau leptons too.

We note that the gauge interactions do not give rise to Ultra--Violet (UV) divergences in the calculation of the Higgs potential, on the contrary,  the fermion Lagrangian defined above provides logarithmic UV divergences which can be removed by suitable conditions among the parameters.
In the $N=2$ case, we can easily derive the conditions for cancellation of the UV divergences in the CW potential (see \cite{BigPaper}) and, among all possible solutions, the enforcement of the ``left-right symmetry'' defined in Ref.~\cite{4dchm} provides the following simple setup, which will be adopted in our analysis:
\begin{align}
&\Delta_L^2 = \Delta_R^1 = 0, \quad Y_2^{11} = M_\Psi^{11},\quad M_\Psi^{21}=0, \notag\\
&Y_1^{11} = Y_1^{22} = Y_1^{21} = Y_2^{21} = Y_2^{22} = 0.
\label{UV}
\end{align}

The low energy Lagrangian for the quark fields, obtained from Eq.~(\ref{ferm}) after the integration of the heavy degrees of freedom, introduces, in general, Flavour Changing Neutral Currents (FCNCs) at tree level via Higgs boson exchanges.
There are basically two ways to avoid such FCNCs. 
The first one is to impose a $C_2$ symmetry~\cite{Mrazek}, just like in $Z_2$ symmetric Elementary 2HDMs (E2HDMs), which would forbid the $Y_1$ term.
The $C_2$ symmetric scenario exactly reproduces a composite realisation of the inert 2HDM. 
In this case the couplings to the SM fields of the SM-like Higgs boson, arising from the $C_2$-even doublet, are the same as in the minimal composite Higgs model \cite{4dchm}, while 
the lightest component of the $C_2$-odd doublet could potentially account for a dark matter candidate.
Here we will follow an alternative approach with a broken $C_2$ which requires an alignment between the two matrices $Y_1$ and $ Y_2$.
Under this assumption, the latter two, for each type of fermions in the low energy Lagrangian, become proportional to each other, like  in the Aligned 2HDM (A2HDM)~\cite{a2hdm}. 

Let us now describe the main steps of the calculation of the Higgs potential.
Integrating out the heavy spin-1 ($\rho^A$ and $\rho_X$) and 1/2 ($\Psi^I$) states, 
we obtain the effective low-energy Lagrangian for the SM gauge bosons, the SM quark fields and the Higgs fields $\Phi_i$. 
In each coefficient of the Lagrangian terms,  form factors appear, which are expressed as functions of the parameters of the strong sector. 
Their explicit forms are provided in \cite{BigPaper}. 
The Higgs potential is then generated  via the CW mechanism by the gauge boson ($W_\mu^a$ and $B_\mu$) 
and  fermion ($q_L$ and $t_R$) loop contributions.  
As  already mentioned, UV divergences do not appear by virtue of the UV-finiteness conditions given in Eq.~(\ref{UV}). 
By expanding  up to the fourth order in the  $\Phi_i$  fields we get
\begin{align}
iV \simeq \frac{1}{f^4}\int\frac{d^4k}{(2\pi)^4}\left[\frac{3}{2}V_G(\Phi_1, \Phi_2)
-6V_F(\Phi_1, \Phi_2)\right],   \label{pot}
\end{align}
 where
$V_{G,F}(\Phi_1, \Phi_2)$ are characterised by the same structure of the Higgs potential in E2HDMs written in terms of 3 dimensionful mass parameters ($m_{11}^2$, $m_{22}^2$ and $m_{12}^2$)
and 7 dimensionless couplings $\lambda_i$ ($i=1,\dots,7$) (see, e.g., Ref.~\cite{Branco} for the definition of these coefficients). 
In general, $m_{12}^2$ and $\lambda_{5,6,7}$ can be complex, but they are found to be real as a consequence of the requirement of CP-conservation in the strong sector. 
These coefficients are determined in terms of the parameters of the strong sector. Therefore, the masses of the Higgs bosons and the scalar mixing angle are fully predicted by the strong dynamics.

{\it Results} --
For our numerical analysis, we take $f_1 = f_2$, $g_{\rho} = g_{\rho_X}$ and $M_{\Psi}^{11} = M_{\Psi}^{22} = M_\Psi^{}$. Then, we have 8 free parameters of the strong sector, i.e.,  
\begin{align}
&f,~g_\rho,~Y_1^{12},~Y_2^{12},~\Delta_L^1,~\Delta_R^2,~M_\Psi^{},~M_{\Psi}^{12}. \label{input}
\end{align}
In order to have phenomenologically acceptable configurations, other than ensuring  Electro-Weak Symmetry Breaking (EWSB), with EW parameters consistent with data, we further require: 
(i) the vanishing of the two tadpoles of the CP-even Higgs bosons, (ii) the predicted top mass to be
$165~\textrm{GeV} < m_t < 175$ GeV and (iii) the predicted Higgs boson mass to be $120~\textrm{GeV} <m_h<130$ GeV. 
Under these constraints, we scan the parameters shown in Eq.~(\ref{input}) within the ranges $0 \leq X \leq 10 f$ ($X = Y_1^{12},~Y_2^{12},~\Delta_L^1,~\Delta_R^2,~M_\Psi^{},~M_{\Psi}^{12}$), 
$600$ GeV $\leq f\leq 3000$ GeV. Hereafter, $g_\rho$ is fixed to be 5. 
As outputs, we obtain the masses of the charged Higgs boson $(m_{H^\pm})$, the CP-odd Higgs boson ($m_A$), the heavier CP-even Higgs boson ($m_H$) and 
the mixing angle $\theta$ between the two CP-even Higgs boson states ($h,H$). Equipped with the mass and coupling spectrum of the C2HDM, we have also tested its parameter space against experimental 
Higgs boson data using {\tt HiggsBounds-5} and {\tt HiggsSignals-2}~\cite{Bechtle:2013wla,Bechtle:2013xfa}.

We highlight next the main differences between our C2HDM and the Minimal Supersymmetric SM (MSSM), 
both of which can be regarded as the minimal realisations of EWSB based on a 2HDM structure embedded in Compositeness and Supersymmetry, respectively. (In the MSSM, the latter is a Type-II one).
For the MSSM predictions, we employ {\tt FeynHiggs 2.14.1} \cite{Heinemeyer:1998yj,Hahn:2013ria} and scan the parameter space according to the recommendations provided in \cite{Bagnaschi:2015hka}.

\begin{figure}[!t]
 \begin{center}
 \includegraphics[scale=0.45]{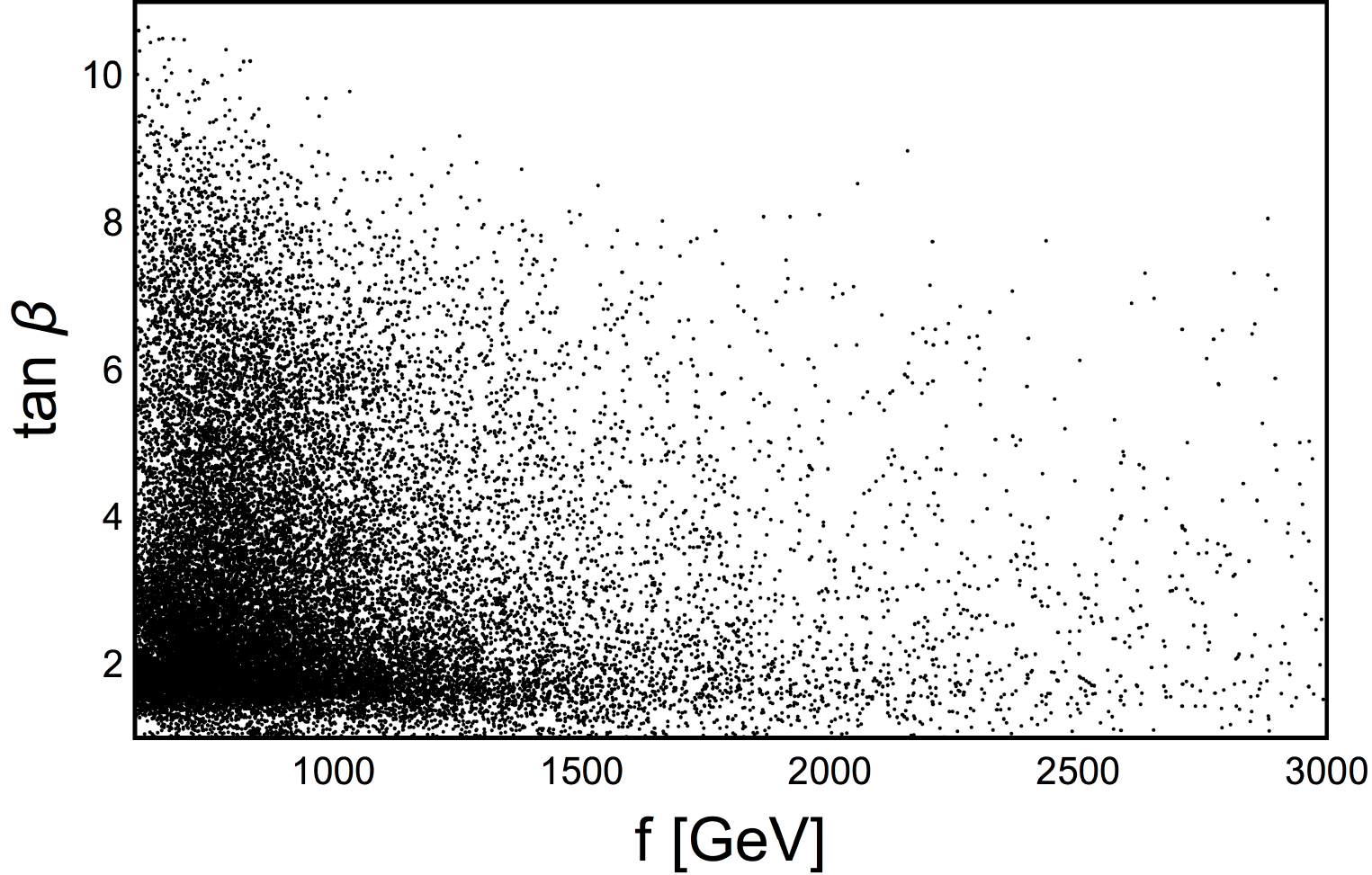}
 \caption{Prediction of $f$ and $\tan\beta$ in the C2HDM.  \label{fig:f-tanB}}
 \end{center}
 \end{figure}

\noindent
\underline{1. Prediction of $\tan\beta$ and  Higgs Boson Masses} --
While in the MSSM the parameter $\tan\beta$ is essentially a free one, albeit potentially limited by theoretically and experimentally constraints, 
in the C2HDM it is predicted and correlates strongly to $f$. This is illustrated in Fig.~\ref{fig:f-tanB}.
We note that all scan points are randomly generated, so that their density is a measure of probability of a region of parameter space to meet the above constraints. 
Clearly, it is seen that the density of the allowed points become smaller in regions with larger values of $f$ and/or $\tan\beta$.
This can be understood by the fact that 
departure from $f \sim v_\textrm{SM}$ requires fine-tuning among the strong parameters, in order to satisfy the tadpole conditions and reconstruct the observed $m_h$ and $m_{t}$ values. 
In fact, this behaviour has also been known in minimal composite Higgs models, see e.g.,~\cite{Panico:2012uw}.
Therefore, in the C2HDM, small $f$ (well within the LHC energy domain)
and $\tan\beta \sim {\cal O}(1)$ (indeed, solutions above $\tan \beta \sim 10$ are highly disfavoured by requiring $m_h \sim 125$ GeV and $m_t \sim 170$ GeV) are naturally predicted. Further, for any $f$ value, we notice that $\tan\beta$ values between (somewhat above) 1 and 6 are more favoured than others. Hence, in the following, we  will at times single out this region of parameter space.

However, this result does not imply that the parameter space of the Higgs sector of the C2HDM is reduced with respect to that of the MSSM, where $\tan\beta$ can in general take values between 1 and, say, $\bar{m}_{t}/\bar{m}_{b}\approx 45$ (with $\bar{m}_{{b,t}}$ being the running masses of the $b,t$-quarks computed at $m_h$), compatible with Supersymmetry unification conditions other than compliant with theoretical and experimental constraints. In fact,  
it should be recalled that $\tan\beta$ is not, in general, a fundamental parameter of a 2HDM, as explained in \cite{Davidson:2005cw,Haber:2006ue,Haber:2010bw}, since it is not basis-independent
and a one-to-one comparison of models for fixed values of $\tan \beta$ is not meaningful unless the realisation of the 2HDM is the same, namely, the models share the same discrete symmetries.
While the MSSM is characterised by a Type-II 2HDM structure, with $\tan \beta$ defined in the basis where the discrete symmetry of the two Higgs doublets is manifest, the C2HDM considered in this work does not possess a $C_2$ symmetry. Even though the strong sector uniquely identifies a special basis for the Higges and, thus, selects a special $\tan \beta$ among all possible basis-dependent definitions 
(see \cite{BigPaper} for more details), this parameter cannot be directly compared to the MSSM one.
Therefore, when comparing physical observables in the composite and  supersymmetric scenarios, one should inclusively span $\tan \beta$ between 1 and 45 for the MSSM and over all predicted values (see Fig.~\ref{fig:f-tanB}) for the C2HDM.

\begin{figure}[!t]
\begin{center}
\includegraphics[scale=0.4]{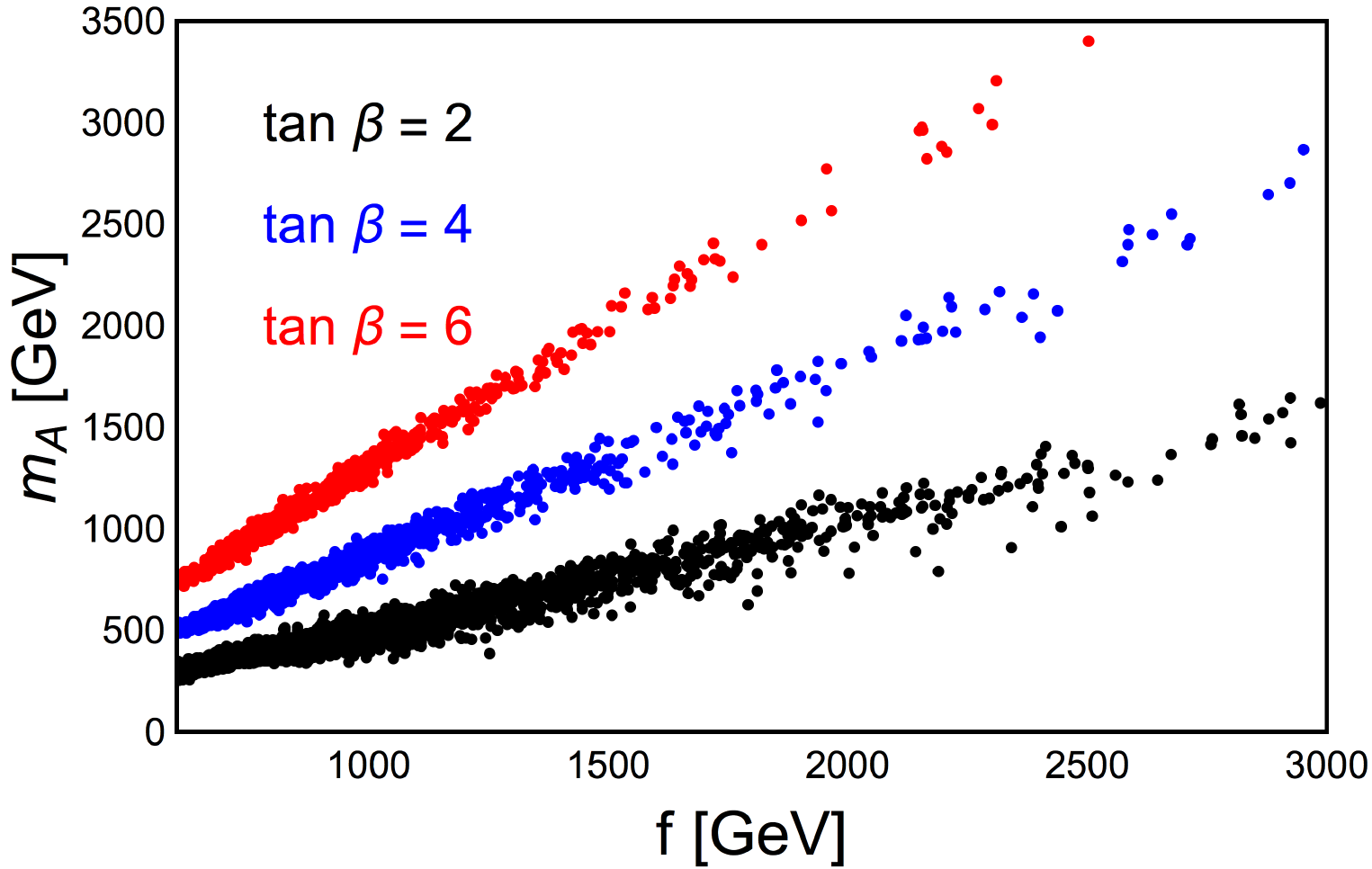}
\caption{Prediction of $m_A^{}$ as a function of $f$ in the C2HDM for $\tan\beta =2$, 4 and 6.  \label{fig:MA}}
\end{center}
\begin{center}
\includegraphics[scale=0.4]{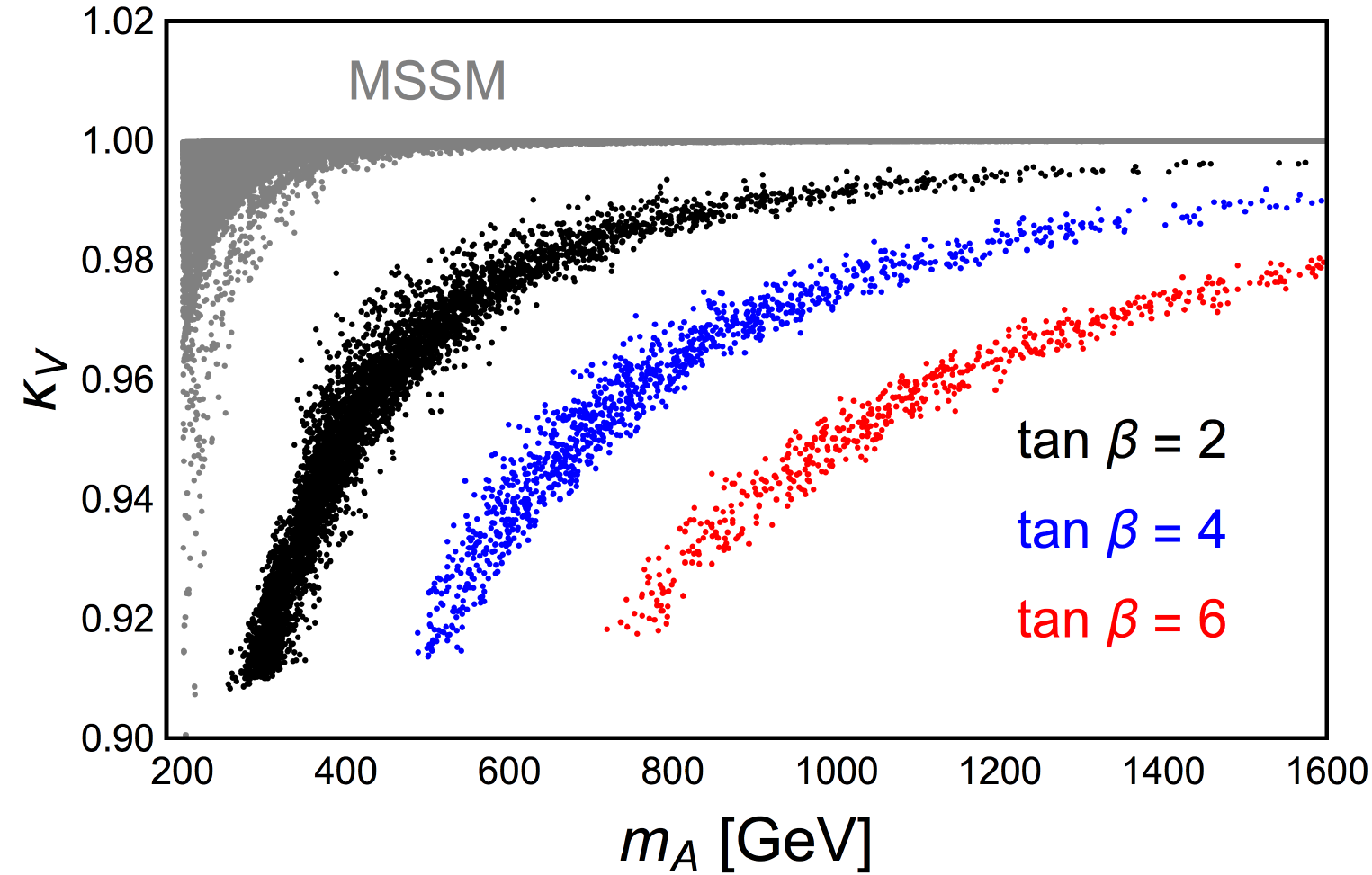} 
\caption{Comparison in the correlation of $m_A^{}$ versus $\kappa_V$ between the C2HDM and  MSSM  
for $\tan\beta=2,4$ and 6 in the former and all values of $\tan\beta$ in the latter.}
\label{fig:DelayedDecoupling}
\end{center}
\end{figure}

Other than $\tan\beta$, also the Higgs masses are  predicted in the C2HDM, e.g., $m_A^{}$ is shown in Fig.~\ref{fig:MA}. 
In the MSSM, in contrast, $m_A^{}$ is normally taken, together with $\tan\beta$, as input value to uniquely define the MSSM Higgs sector at tree level 
(although now the discovered SM-like Higgs boson, identified  with the $h$ state, removes the arbitrariness of the $m_A$ choice, at least at lowest order). In particular, we find that larger values of $m_A^{}$ are obtained for larger $f$ and/or $\tan\beta$. 
The mass of the pseudoscalar is not directly constrained by the tadpole conditions and it is naturally of order $f$. In particular, one can show that 
$m_A^2 \simeq c (1 + \tan^2 \beta) f^2$,
where $c \sim 0.05$ may vary by a factor of 2 only for $\tan \beta \lesssim 1$.
All these features remain stable against different choices of $g_\rho\, > 1$.

\noindent
\underline{2. Alignment with Delayed Decoupling} --  In addition to Higgs masses, further physics observables that can be used to compare the C2HDM to the MSSM are, e.g., Higgs cross sections and branching ratios. A convenient way to exploit 
the latter in order  
to extract the model parameters of potential new physics in the Higgs sector   is to recast them in
 the  language of the so called $\kappa_i$ `modifiers'  of Ref. 
\cite{LHCHiggsCrossSectionWorkingGroup:2012nn}, wherein any of the latter is nothing but a coupling of the SM-like Higgs boson discovered at the LHC to known  fermions ($i=b,t,\tau$) and bosons ($i=g, \gamma,Z,W^\pm$) normalised to the corresponding SM prediction. In order to compare the C2HDM and MSSM in this framework, we turn now to the case of $\kappa_V$ ($V=Z,W^\pm$), this being the most precisely known of all $\kappa_i$'s. 
In the C2HDM the $hVV$ ($V=Z,W^\pm$) coupling, normalised to the SM prediction, is given by  
\begin{equation}
\kappa_V = \left(1-\frac{\xi}{2} \right)\cos\theta, \quad \xi \equiv \frac{v_\textrm{SM}^2}{f^2}, 
\label{eq:kappaV}
\end{equation}
where $\theta \to 0$ with $f \to \infty$ corresponds to the alignment limit, i.e., the couplings of $h$ to SM particles become the same as those of the SM Higgs boson at tree level.  
Fig.~\ref{fig:DelayedDecoupling} shows that the (near) alignment limit ($\kappa_V\sim 1$) is reached at large Higgs boson masses 
(exemplified here by the CP-odd one) in both the C2HDM and MSSM. 
However, a remarkable difference is seen in this behaviour. In the MSSM, $\kappa_V^{}$ (see \cite{Djouadi} for its expression) 
very quickly reaches~1.   
In contrast, in the C2HDM, $\kappa_V$  approaches 1 slowly (and the velocity at which this happens depends strongly on $\tan\beta$). 
This delayed decoupling is mainly driven by the negative ${\cal O}(\xi)$ corrections which are typical of composite Higgs models, as seen in Eq.~(\ref{eq:kappaV}), combined with the fact that  the slopes seen in Fig.~\ref{fig:DelayedDecoupling} for the C2HDM exhibit the dependence of $m_A$ from $f$ illustrated in Fig.~\ref{fig:MA},  
which allows a different and wider spread of $\kappa_V$ values away from 1 with respect to the MSSM. 
We note that values of $\kappa_V \gtrsim 0.9$ are currently compatible with LHC data at 1$\sigma$  level~\cite{Khachatryan:2016vau}. 
Therefore, if a large deviation in the $hVV$ coupling from the SM prediction will be restricted by future experiments, 
it will imply a larger Compositeness scale in the C2HDM. Conversely, if such a deviation will  instead be established and no heavy Higgs state below 400 GeV or so will be seen, the MSSM may be ruled out and the C2HDM could explain the data. 
Therefore, either way, by combining the measured value of $\kappa_V^{}$ to that of an extracted or excluded $m_A$, one may be able to discriminate the C2HDM from the MSSM. 

\begin{figure}[h]
\begin{center}
\includegraphics[scale=0.6]{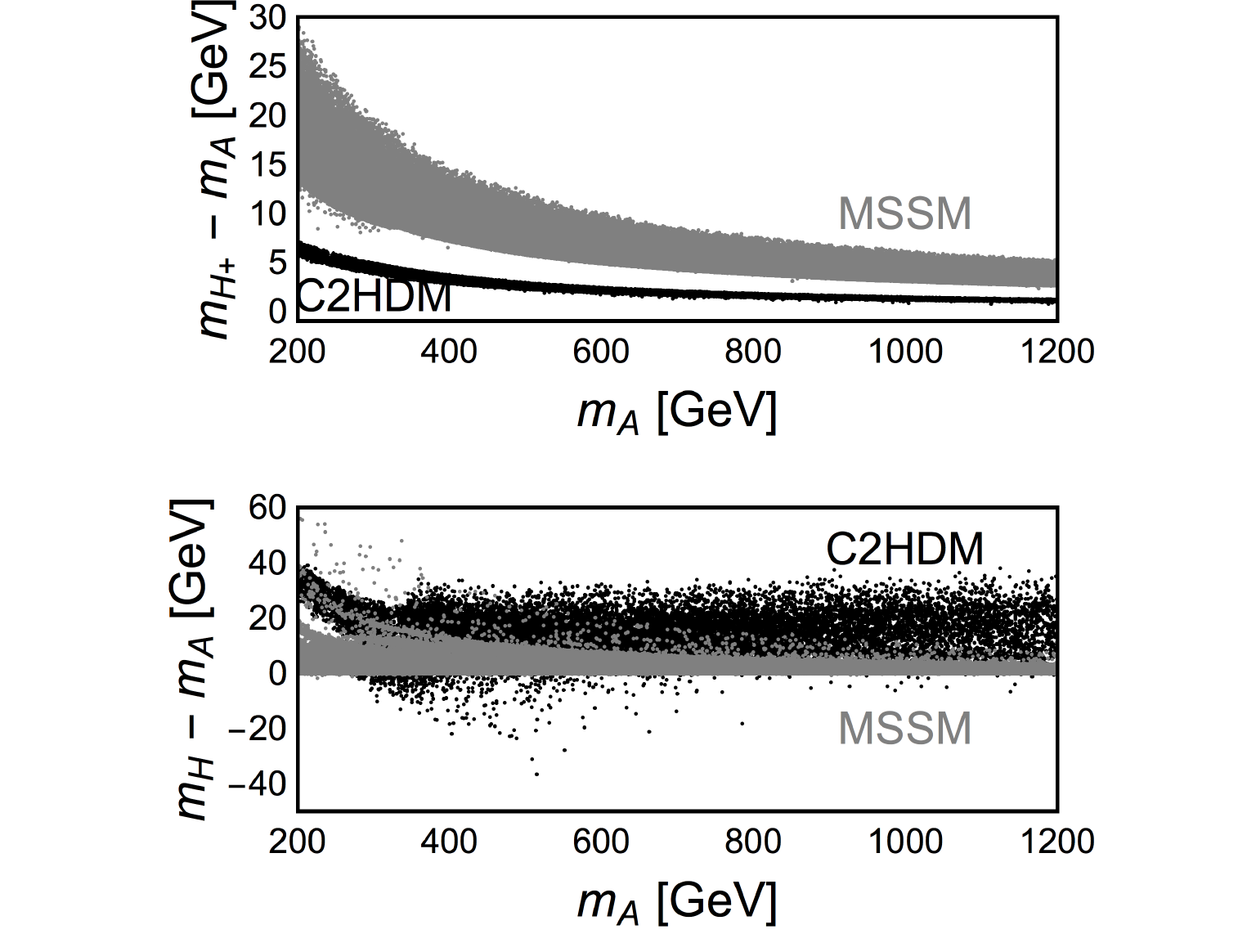}
\caption{Comparison in the correlation of $m_A$ versus $m_{H^\pm}^{} - m_A^{}$ (top) and $m_H^{} - m_A^{}$ (bottom) between the C2HDM and  MSSM for all values of $\tan\beta$ in both scenarios. }
\label{fig:NonDegeneracy1}
\end{center}
\end{figure}

\noindent
\underline{3. Mass Hierarchy Amongst Heavy Higgs States} --
Fig.~\ref{fig:NonDegeneracy1} shows the mass differences $m_{H^{\pm}}^{}-m_A^{}$ and $m_{H}^{}-m_A^{}$ where $\tan\beta$ has been varied over all its possible values in our two reference models, as explained above.
We find that (top frame), while $m_{H^\pm}$ and $m_A$ are very close in the C2HDM, within 5 GeV, 
larger mass differences between these two heavy Higgs bosons are allowed in the MSSM,  particularly for smaller  $m_A$, e.g.,   $m_{H^\pm}^{} - m_A^{}$ can reach $\approx 30$ GeV for $m_A=200$ GeV. 
Due to the particular structure of the scalar potential, the contribution of the fermionic sector cancels out in the mass splitting of $A$ and $H^\pm$ and only the gauge sector one survives. The latter can be approximated by $(m_{H^\pm} - m_A)/m_A \simeq g'^2 \xi$,
with $g'$ the hypercharge gauge coupling. This represents a robust prediction of the model.
Conversely, for $m_{H}^{} - m_A^{}$  (bottom frame), it is the other way around. With increasing $m_A$, starting from 300 GeV, the mass difference between
 the two heavy neutral Higgs bosons tends to be confined within 5 GeV or so for the MSSM while  in the C2HDM this can range from $-40$ GeV (at moderate $m_A$) to $+40$ GeV (for larger $m_A$).
The mass splitting is not strictly determined as in the $m_{H^\pm} - m_A$ case but can be, nevertheless, estimated by $(m_H - m_A)/m_A \simeq c \, \xi$ with $c$ being an order 0.1 coefficient encoding the dependence on the fermionic parameters of the strong sector. The $\xi$ factor, appearing in the formulas of the mass splittings, reproduces the expected mass degeneracy among $A, H$ and $H^\pm$ in the large $f$ regime.
Interestingly then, the hierarchy amongst $m_{H^\pm}$, $m_A$ and $ m_H$ may enable one to distinguish between the two scenarios as (recalling that three-body decays via off-shell gauge bosons are possible) establishing  $H^\pm\to W^{\pm *} A$ would point to the MSSM while extracting $H\to Z^*A$ or $A\to Z^* H$ would favour the C2HDM.

\begin{figure}[h]
 \begin{center}
 \includegraphics[scale=0.4]{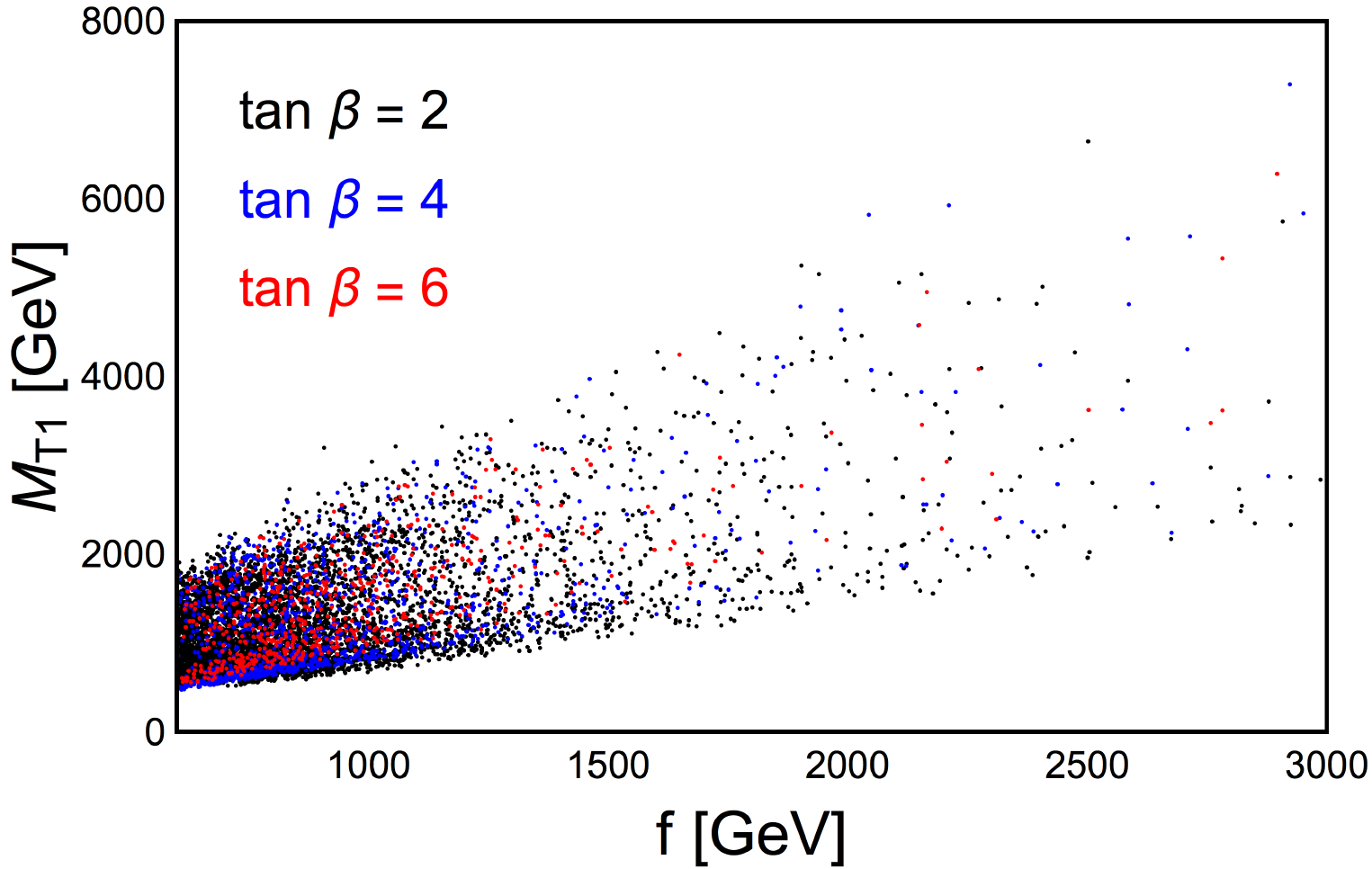} 
 \caption{Correlation of $M_{T_1}$ versus $f$ in the C2HDM for $\tan\beta = 2$, 4 and 6.}
 \label{fig:HeavyFermions1}
 \end{center}
 \begin{center}
\includegraphics[scale=0.27]{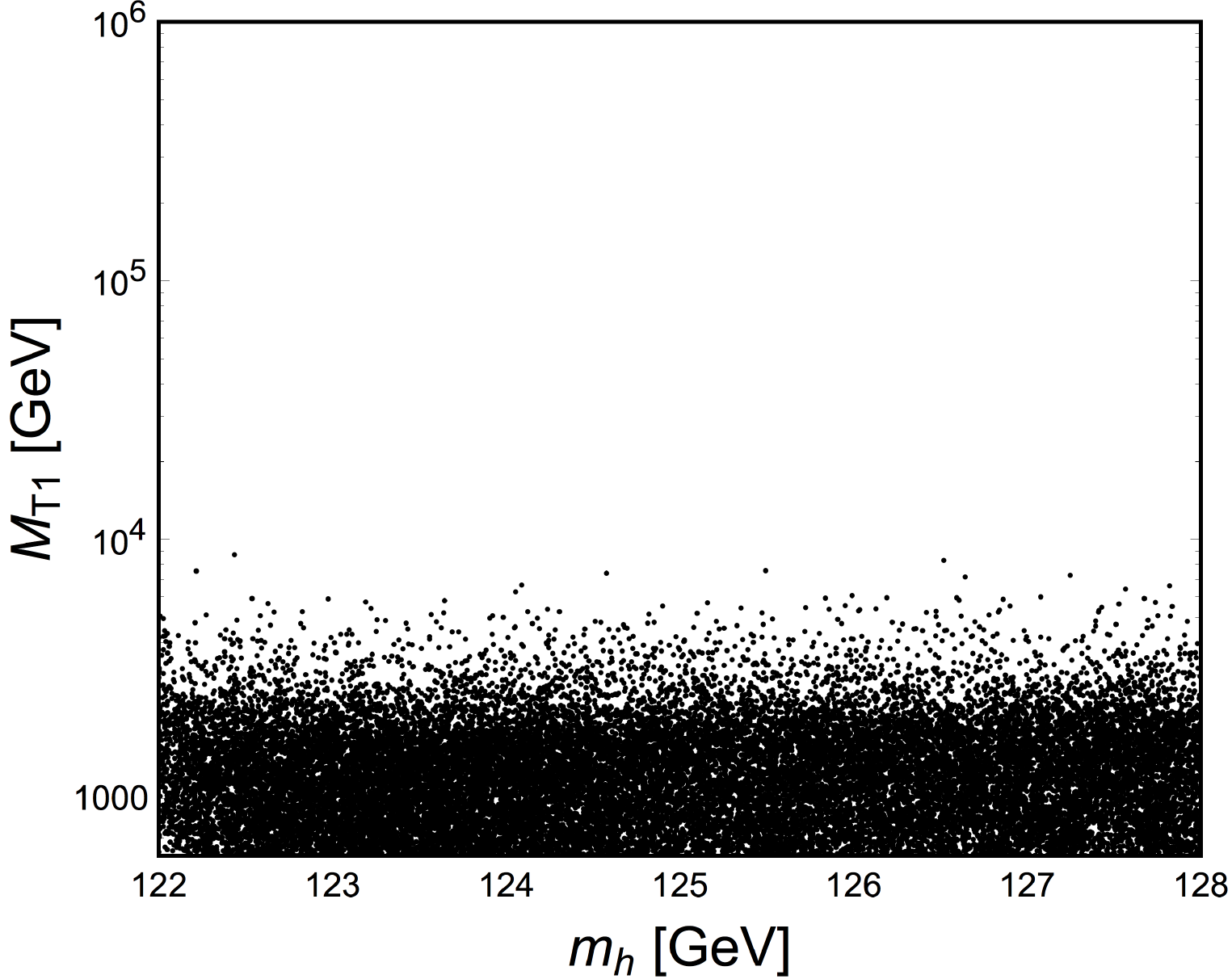} 
 \includegraphics[scale=0.27]{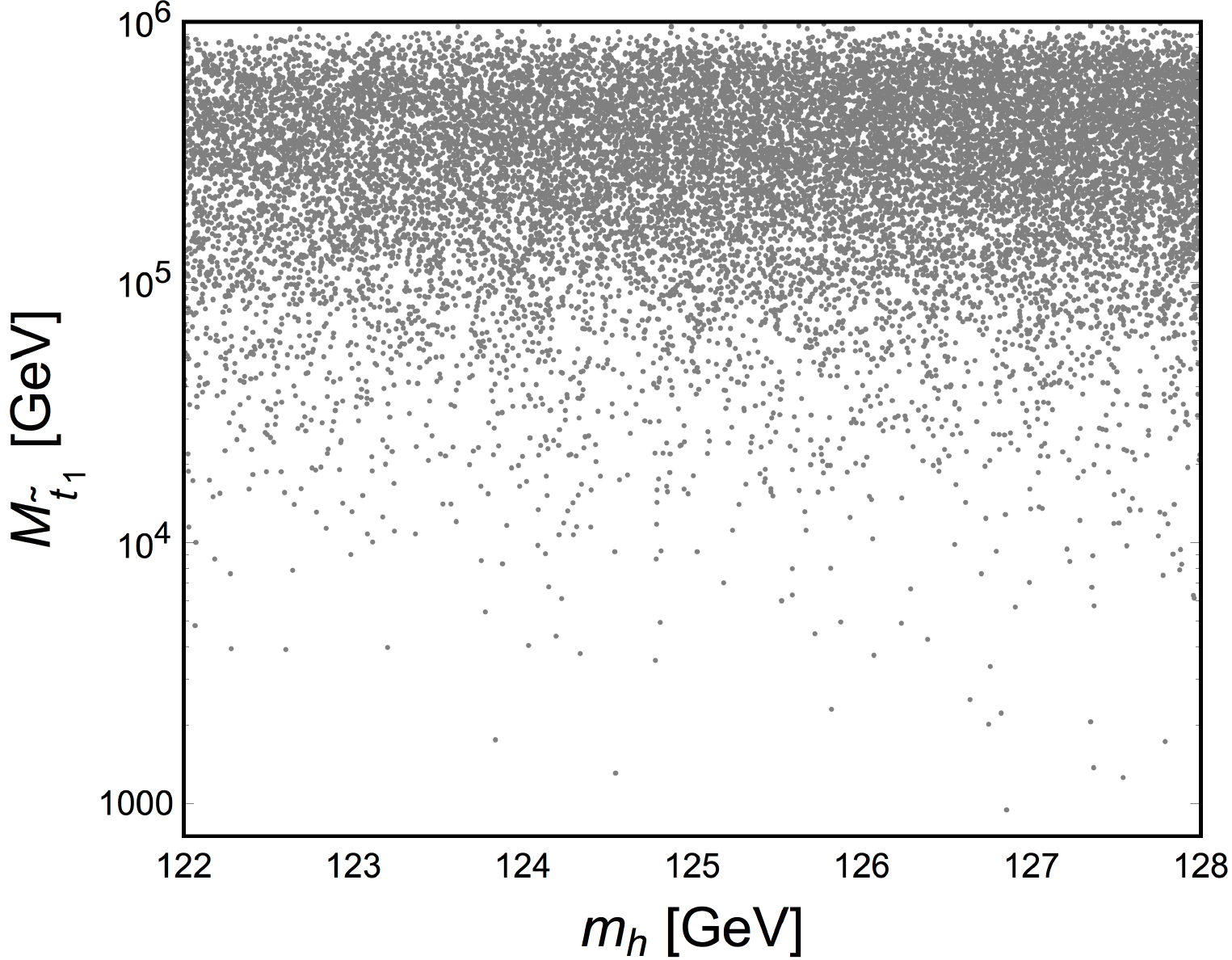} 
 \caption{Left: Correlation of $m_h$ versus $M_{T_1}$, the mass of the lightest 2/3 fermionic partner in the C2HDM.   
Right: Correlation of $m_h$ versus $M_{{\tilde t}_1}$, the mass of the lightest stop in the MSSM. Plots are 
for all values of $\tan\beta$ in both scenarios.
}
 \label{fig:HeavyFermions2}
 \end{center}
 \end{figure}

\noindent
\underline{4. Lower-lying Top Quark Partners} --
Needless to say, just like the Higgs masses, the composite top quark partner masses are strongly correlated to  $f$. 
In Fig.~\ref{fig:HeavyFermions1}, we show, e.g.,  the correlation between $f$ and the lightest top partner mass $M_{T_1}^{}$. Here, we can see that the $\tan\beta$
dependence is rather unimportant.  
In particular, we find that typically a minimum $M_{T_1}$  is required and such a lower limit  gets higher as $f$ increases. 
For a given $f$, the minimum allowed value of $M_{T_1}$ is $\sim  f$, and it is strictly determined by the reconstruction of the top mass from the parameters of the strong sector.
This behaviour agrees with well-established results in minimal composite Higgs models \cite{4dchm, Panico:2012uw}.
A distinctive feature between the C2HDM and MSSM in connection with the heavy (s)top sector is the fact that the measured value of $m_h\sim 125$ GeV
requires a (fermionic) top quark partner in the C2HDM with a mass significantly lower than the (scalar) top quark partner in the
MSSM.
The lightest $2/3$ fermion partner in the C2HDM
is then potentially accessible at the LHC whatever the actual $m_h$ value, see Fig.~\ref{fig:HeavyFermions2}.
This is much unlike the  MSSM, for which the stop mass is, over the majority of the parameter space, far beyond the reach of the LHC \cite{Djouadi:2013vqa,Djouadi:2015jea}.    
In essence, present knowledge of the Higgs sector implies that the C2HDM is more readily accessible in the top partner sector than the MSSM. It is worth noticing that this result relies on the particular structure of the superpotential of the MSSM and could be modified in other supersymmetric scenarios. An interesting example is provided by the Next-to-MSSM (NMSSM) in which the presence of a SM-singlet chiral superfield allows for more freedom in the calculation of the Higgs mass. This implies that the dependence of $m_h$ on the lightest stop mass is relaxed with respect to the MSSM so that $M_{\tilde t_1}$ could be lowered down to the TeV region. Likewise, Compositeness models with different global symmetries and/or fermionic representations may result in different Higgs and/or top partner spectra.

{\it Conclusions} --
We have calculated the mass and coupling spectra of a C2HDM based on the global symmetry breaking of $SO(6) \to SO(4)\times SO(2)$, as all these physical quantities are predicted by the dynamics of such a strong sector. 
In particular, we have focused on the differences between predictions given in the C2HDM and MSSM, both of which provide a 2HDM as a low energy effective theory of EWSB, specifically, an A2HDM in the case of Compositeness  and a Type-II 2HDM in the case of Supersymmetry.  
We found remarkable differences between the C2HDM and  MSSM. 
Namely, (i) $\tan\beta$ is predicted in the C2HDM, whereas this parameter is arbitrary in the MSSM, 
(ii) all  Higgs boson masses are also predicted within the C2HDM for a given value of $f$ (unlike in the MSSM), 
(iii) the speed of  decoupling of the extra Higgs bosons in the C2HDM is much slower than in the MSSM, 
(iv) different size mass splittings amongst the extra neutral Higgs states are predicted in the C2HDM with respect to the MSSM, so that establishing different Higgs-to-Higgs plus gauge boson decays  at the LHC  could potentially distinguish
between the two scenarios, 
(v) the mass of the lightest (fermionic) top quark partner in the C2HDM can be much smaller than the lightest (scalar) stop mass in the MSSM.
Remarkably, all these aspects are
amenable to investigation at current LHC experiments in the years to come, 
so that a clear potential exists in the foreseeable future to disentangle two possible solutions of the SM hierarchy problem, i.e., Compositeness and Supersymmetry. 

{\it Acknowledgements} --
The work of  SM is supported in part by the NExT Institute and the STFC Consolidated Grant 
ST/L000296/1. We all thank Sven Heinemeyer, Suchita Kulkarni and Andrea Tesi for illuminating discussions.


\begin{thebibliography}{1}
\bibitem{BigPaper} 
S.~De Curtis, L.~Delle Rose, S.~Moretti, A.~Tesi and K.~Yagyu, in preparation.

 
 \bibitem{Djouadi}
  A.~Djouadi,
  Phys.\ Rept.\  {\bf 459} (2008) 1. 
 
\bibitem{partial}
  D.~B.~Kaplan and H.~Georgi,
  Phys.\ Lett.\  B {\bf 136} (1984) 183.


\bibitem{4dchm} 

  S.~De Curtis, M.~Redi and A.~Tesi,
  JHEP {\bf 1204} (2012)  042.



 
\bibitem{Mrazek} 
  J.~Mrazek, A.~Pomarol, R.~Rattazzi, M.~Redi, J.~Serra and A.~Wulzer,
  Nucl.\ Phys.\ B {\bf 853}  (2011) 1. 

\bibitem{a2hdm} 
  A.~Pich and P.~Tuzon,
  Phys.\ Rev.\ D {\bf 80} (2009) 091702. 


\bibitem{Branco} 
  G.~C.~Branco, P.~M.~Ferreira, L.~Lavoura, M.~N.~Rebelo, M.~Sher and J.~P.~Silva,
  Phys.\ Rept.\  {\bf 516} (2012) 1. 

\bibitem{Bechtle:2013wla}
  P.~Bechtle, O.~Brein, S.~Heinemeyer, O.~Stål, T.~Stefaniak, G.~Weiglein and K.~E.~Williams,
  Eur.\ Phys.\ J.\ C {\bf 74} (2014) no.3,  2693
  
\bibitem{Bechtle:2013xfa}
  P.~Bechtle, S.~Heinemeyer, O.~Stål, T.~Stefaniak and G.~Weiglein,
  Eur.\ Phys.\ J.\ C {\bf 74} (2014) no.2,  2711
  
\bibitem{Heinemeyer:1998yj}
  S.~Heinemeyer, W.~Hollik and G.~Weiglein,
  Comput.\ Phys.\ Commun.\  {\bf 124} (2000) 76.

\bibitem{Hahn:2013ria}
  T.~Hahn, S.~Heinemeyer, W.~Hollik, H.~Rzehak and G.~Weiglein,
  Phys.\ Rev.\ Lett.\  {\bf 112} (2014)  141801.

\bibitem{Bagnaschi:2015hka}
  E.~Bagnaschi {\it et al.},
  LHCHXSWG-2015-002.

\bibitem{Panico:2012uw}
  G.~Panico, M.~Redi, A.~Tesi and A.~Wulzer,
  JHEP {\bf 1303} (2013) 051. 

\bibitem{Davidson:2005cw}
  S.~Davidson and H.~E.~Haber,
  Phys.\ Rev.\ D {\bf 72} (2005) 035004
   [Erratum: Phys.\ Rev.\ D {\bf 72} (2005) 099902].

\bibitem{Haber:2006ue}
  H.~E.~Haber and D.~O'Neil,
  Phys.\ Rev.\ D {\bf 74} (2006) 015018
   [Erratum: Phys.\ Rev.\ D {\bf 74} (2006)  059905].

\bibitem{Haber:2010bw}
  H.~E.~Haber and D.~O'Neil,
  Phys.\ Rev.\ D {\bf 83} (2011) 055017.

\bibitem{LHCHiggsCrossSectionWorkingGroup:2012nn}
  A.~David {\it et al.} [LHC Higgs Cross Section Working Group],
  arXiv:1209.0040 [hep-ph].
%

\bibitem{Khachatryan:2016vau}
  G.~Aad {\it et al.} [ATLAS and CMS Collaborations],
  JHEP {\bf 1608} (2016) 045. 

\bibitem{Djouadi:2013vqa}
  A.~Djouadi and J.~Quevillon,
  JHEP {\bf 1310} (2013) 028.

\bibitem{Djouadi:2015jea}
  A.~Djouadi, L.~Maiani, A.~Polosa, J.~Quevillon and V.~Riquer,
  JHEP {\bf 1506} (2015) 168.




\end{thebibliography}
\end{document}